%% file: PE26_10_02.tex
\documentclass[twocolumn,amsmath,floatfix,prl,aps,superscriptaddress]{revtex4-2}
\usepackage{graphicx}
\usepackage{booktabs}
\usepackage[caption=false]{subfig}
\usepackage{amsmath, amssymb}
\usepackage{hyperref}

\input{dft}

\begin{document}

\title{Electronic Density versus Geometry for Machine-Learned Molecular Absorption Spectra}

\author{S. Dhanpal}
\affiliation{Scientific Computing Department, Science and Technology Facilities Council, Rutherford Appleton Laboratory, Didcot OX11 0QX, United Kingdom.}
\author{P. Elliott}
\affiliation{Scientific Computing Department, Science and Technology Facilities Council, Rutherford Appleton Laboratory, Didcot OX11 0QX, United Kingdom.}
\author{P. E. Trevisanutto}
\affiliation{Scientific Computing Department, Science and Technology Facilities Council, Daresbury Laboratory, Warrington, WA4 4AD, United Kingdom.}
\author{A. M. Elena}
\affiliation{Scientific Computing Department, Science and Technology Facilities Council, Daresbury Laboratory, Warrington, WA4 4AD, United Kingdom.}
\author{G. Teobaldi}
\affiliation{Scientific Computing Department, Science and Technology Facilities Council, Rutherford Appleton Laboratory, Didcot OX11 0QX, United Kingdom.}

\date{\today}

\begin{abstract}
Molecular optical absorption spectroscopy provides a direct probe of electronic structure and is widely used for molecular identification, interpretation of photophysical behaviour, and planning of spectroscopy experiments. 
Calculating the absorption spectra using first-principle excited-state methods, however, is computationally demanding, at least compared to ground-state calculations, which limits their routine application across large molecular sets. 
Machine-learning (ML) surrogates can reduce this cost and allow rapid spectral prediction, which can help laboratory users screen candidate molecules, select experimentally relevant energy windows, and compare alternative molecular assignments. However, their performance depends strongly on how molecular information is represented. 
Here, we compare using the ground-state electron density versus the molecular geometry as inputs to a ML model for predicting absorption spectra, for a training set of 6874 molecules selected from the QM7 dataset. For each of these molecules, the density was calculated using density functional theory (DFT) and the absorption spectrum was calculated using linear-response (LR) time-dependent DFT (TDDFT). 
Utilizing the ground-state density as the input to the ML model is motivated by the Hohenberg-Kohn and Runge-Gross theorems, and the fact that the ground-state density encodes information about bonding, charge localisation, and electronic delocalisation. Hence, it may be a more judicious starting point for the ML model compared to the geometry, as it effectively decouples the chemistry of the ground-state.
The question we test is whether the benefits of using the density outweigh the (notprohibitive) penalty of requiring an additional single-point DFT calculation for the density. 
We find that the density-based convolutional neural network achieves a validation correlation of 0.9926, compared with 0.9795 for the best geometry-based graph model, reducing the residual decorrelation, by approximately $64\%$.
\end{abstract}
\maketitle

\section*{Introduction}

The interaction of light with molecules is one of the most direct probes of electronic structure. Optical absorption spectra encode the excitation energies, oscillator strengths, charge redistribution, and selection rules associated with the response of the electronic system to an external electromagnetic perturbation. In this sense, an electronic absorption spectrum is not only a fingerprint for chemical identification, but also a compact representation of bonding, conjugation, charge-transfer character, frontier-orbital structure, and the local electronic environment. Reliable computation and interpretation of such spectra is therefore central to molecular spectroscopy, photochemistry, materials discovery, and the design of chromophores, dyes, emitters, sensors, and photoactive molecules.

The standard first-principles route to molecular absorption spectra is based on density-functional workflows, where one first solves a ground-state Kohn--Sham density-functional theory (DFT) problem, obtaining orbitals, eigenvalues, and the ground-state electron density \cite{hohenberg1964,kohn1965}. Excited-state information is then obtained through a linear-response (LR) time-dependent density-functional theory (TDDFT)
\cite{runge1984,GK85,C96,PGG96,onida2002,EBF07,U14}. 
These
calculations yield excitation energies and oscillator strengths, which are typically broadened using a prescribed line-shape function in order to compare to experimental spectra. TDDFT offers a useful compromise between quantum-chemical accuracy and computational cost, and it remains a workhorse method for UV--visible spectroscopy. Nevertheless, despite favorable scaling compared to other excited-state methods, the LR-TDDFT step is often the computational bottleneck: it involves solving a large eigenvalue-eigenvector problem in the space of single-particle transitions (or inverting a large matrix when solving a Dyson-like equation for periodic systems\cite{PGG96,onida2002}). This cost becomes particularly restrictive when many excited states, broad spectral windows, or large numbers of candidate molecules are required \cite{ruger2015,brabec2015}.

Fast computation of spectra and electronic gaps would therefore have immediate practical value. In high-throughput molecular screening, a rapid surrogate can be used to identify candidates with target absorption windows before more expensive TDDFT or higher-level calculations are performed. In experimental spectroscopy, predicted spectra can assist the assignment of peaks, guide the
selection of pump and probe wavelengths, and support the interpretation of gas-phase and condensed-phase measurements. This is especially relevant for large experimental facilities where beam time is limited and where theory is used to plan or interpret measurements \cite{clfArtemis}.

Machine learning has become a powerful route for constructing such surrogates. Early molecular machine-learning work demonstrated that quantum properties of organic molecules could be learned across chemical compound space from nuclear charges and Cartesian coordinates \cite{montavon2013}. The development of benchmark datasets such as MoleculeNet and QM9 further established molecular property prediction as a central task in scientific machine learning  \cite{ramakrishnan2014qm9,wu2018moleculenet}. Modern atomistic neural networks then introduced increasingly expressive representations of molecular geometry. SchNet used continuous-filter convolutions to learn smooth atomistic representations directly from nuclear coordinates \cite{schutt2018schnet}; DimeNet and DimeNet++ incorporated angular information through directional message passing \cite{gasteiger2020dimenet,gasteiger2020dimenetpp}; and equivariant models such as NequIP and MACE showed that enforcing Euclidean symmetry at the architecture level can substantially improve data efficiency and accuracy for atomistic modelling \cite{batzner2022nequip,batatia2022mace}. These models have made geometry-based graph neural networks a natural baseline for molecular quantum-property prediction.

The extension of machine learning from ground-state scalar properties to excited states and spectra\cite{RHTL15,KCR22,BSZX25,SCMI26, GGR26,FSC23,SC26,NS17,GSTJ19,KSL20,XBD20,GGG22,JJC24} is more recent, but rapidly developing. For example, SchNarc combined SchNet-like representations with excited-state dynamics workflows and was subsequently used to predict quantities needed for Vis-UV absorption spectra \cite{westermayr2020schnarc,westermayr2020uv}. Models have also been shown to predict X-ray absorption spectra with near first-principles fidelity \cite{carbone2020xas,PWMD24,CCHD24}. Other work has argued that purely geometric descriptors are often insufficient for spectroscopic prediction and that low-cost electronic descriptors inspired by TDDFT can substantially improve supervised absorption-spectrum models \cite{dearmas2023spectroscopic}. 

This motivates the use of the electron density as an input representation. The electron density has a key status in electronic-structure theory due to the Hohenberg--Kohn theorem, which establishes that the ground-state electron density determines the external potential, up to an additive constant, and hence the electronic Hamiltonian and all ground-state observables \cite{hohenberg1964}. This also includes the ground-state absorption spectrum as all ingredients to the equations of LR-TDDFT depend only on the ground-state density. 

This provides a strong physical motivation for using the ground-state density as an input variable, since its real-space distribution encodes information about the electronic environment from which the optical response emerges, although care must be taken for certain hypersensitive properties\cite{AWLH26}. 

A density-based machine-learning model does not replace quantum chemistry end to end, since the ground-state density must still be computed. The aim is to retain the cheaper ground-state DFT step while replacing the more expensive density-to-response calculation with a learned surrogate. This places the model at a physically meaningful intermediate stage, where spectra are learned from an electronic field rather than only from atom types and nuclear coordinates. This viewpoint is supported by density-centred machine learning, where electron densities have been predicted in condensed-phase systems and used to recover downstream electronic properties \cite{lewis2021salted,RVMJ25}. Related work has shown that density responses can also provide field responses, polarizabilities, and Raman observables \cite{lewis2023salter}. More broadly, OMol25 and response-aware models such as MACE-Field, MACE-MDP, and MACE-POLAR-1 reflect a growing move toward models that include electronic response, long-range electrostatics, polarizability, and density-like information beyond local nuclear geometry \cite{levine2025omol25,goennheimer2026macemdp,martin2025macefield,batatia2026macepolar}.

In this work, we investigate this representation question directly. We study machine-learning models for a related molecular electronic-structure task: prediction of the full broadened absorption spectrum on a fixed frequency grid. We compare two broad classes of representation. The first uses volumetric ground-state electron densities processed by three-dimensional convolutional neural networks. The second uses atom-centred molecular geometries processed by graph neural networks, including SchNet-like, directional message-passing, and equivariant architectures. %

\section*{Methodology}
\label{sec:methodology}

\subsection*{Study overview and prediction tasks}
\label{sec:study_overview}

Starting from the atomic species and molecular geometries provided in the QM7 dataset\cite{rupp2012qm7,montavon2013,YLWG19}, we constructed a derived electronic-structure dataset containing ground-state electron densities, molecular absorption spectra, and several electronic-gap quantities.

In total, $6874$ molecular calculations converged and produced the quantities required for the present study. For each molecule, we consider two complementary input representations. The geometry-based representation consists of the atomic numbers and nuclear coordinates,
\begin{equation}
    \mathcal{G}
    =
    \left\{
        Z_i,\mathbf{R}_i
    \right\}_{i=1}^{N_{\mathrm{at}}},
    \label{eq:geometry_representation}
\end{equation}
where \(Z_i\) and \(\mathbf{R}_i\) denote the nuclear charge and Cartesian
position of atom \(i\), respectively. The density-based representation
consists of the ground-state electron density,
\(\rho(\mathbf{r})\), sampled on a regular three-dimensional Cartesian grid.

The primary prediction task is the molecular absorption spectral shape on a
fixed photon-energy grid. The reference spectra were generated using
time-dependent density-functional theory and were represented by
\(L=900\) frequency bins over the interval
\begin{equation}
    0 \leq \omega \leq 0.45\ \mathrm{a.u.},
    \label{eq:spectral_window}
\end{equation}
corresponding approximately to \(0\)--\(12.2\) eV. This interval spans the ultraviolet and part of the vacuum-ultraviolet region and was selected as the spectral range investigated in the present work. %

Using common reference calculations and identical molecular splits for the two input representations enables a direct comparison of the information provided by nuclear geometry and ground-state electron density. The comparison is not intended solely as an architecture benchmark. Rather, it tests whether the preferred molecular representation depends on the target observable. %

\subsection*{Electronic-structure dataset construction}
\label{sec:dataset_construction}

\subsubsection*{Source molecular structures}
\label{sec:qm7}

The QM7 dataset provides the atomic species and Cartesian coordinates of 7211 small organic molecules \cite{YLWG19,rupp2012qm7,montavon2013}. These geometries were used as the starting point for all subsequent calculations. We select from QM7, $6874$ small organic molecules composed of
H, C, N, and O, with up to seven non-hydrogen atoms . The atomic geometries supplied by QM7 were
used as the common structural input for the electronic-structure calculations
and for the geometry-based machine-learning models. QM7 itself does not provide the volumetric electron densities or LR-TDDFT spectra used as targets in this work; these quantities were generated specifically for the present study.

The molecular orientation inherited from the source geometry was retained consistently during the generation of the ground-state density and the corresponding machine-learning input. No molecule was represented by different orientations in different data subsets.

\subsubsection*{Ground-state density-functional theory}
\label{sec:ground_state_dft}

The ground-state electronic density is obtained by solving the Kohn-Sham (KS) equation of DFT:
\begin{equation}
\left(-\frac{\nabla^2}{2} +v\s(\br) \right) \phi_{j}(\br) = \epsilon_j\phi_{j}(\br) 
\label{e:KS}
\end{equation}
and calculating 
\begin{equation}
\rho(\br) = \sum_j^{N/2} |\phi_j(\br)|^2
\label{e:rho}
\end{equation}
to find the density, where j labels the eigenstates of Eq. (\ref{e:KS}), i.e. the eigenvalue $\epsilon_j$ and the orbital $\phi_{j}(\br)$, and N is the number of electrons in the molecule. For closed-shell molecules, each orbital is doubly occupied, hence the sum in Eq. (\ref{e:rho}) has $N/2$ terms. The KS potential, $v\s(\br) = v\ext(\br) + v\H[\rho](\br) + v\xc[\rho](\br)$,  consists of the external potential $v\ext(\br)$ constructed from the molecular geometry, the Hartree potential $v\H[\rho](\br)=\int d^3r' \rho(\br')/|\br-\br' |$, and the exchange-correlation (XC) potential $v\xc[\rho](\br) = \delta E\xc[\rho]/\delta\rho(\br)$, where $E\xc[\rho]$ is the XC energy density functional which is unknown, and an approximation must be chosen for any practical calculation. 

\subsubsection*{Linear-Response TDDFT absorption spectra}
\label{sec:rttddft}

The absorption spectrum for each molecule is calculated using the so-called Casida equation\cite{casida1995} of linear-response (LR) TDDFT:
\ben
\underline{\underline{\Omega}} = \omega^{2}_{q} \delta_{qq'} + 2 \sqrt{\omega_{q}\omega_{q'}}K_{qq'} 
\label{e:casida}
\een
where matrix $\underline{\underline{\Omega}}$ has eigenvalues $\Omega_q^2$ with $\Omega_q$ corresponding to the excitation frequencies of the molecule. The compound label $q$ is a pair of DFT eigenstates, an occupied level $j$ and an empty level $k$, and so $\omega_q = \epsilon_k-\epsilon_j$. The matrix $K_{qq'}$ is defined as:
\ben
K_{qq'} = \int d\br\int d\br' \ \Phi_{q}(\br)\left(f\H(\br,\br')+f\xc(\br,\br')\right)\Phi^{*}_{q'}(\br') 
\een
where $\Phi_q=\phi_j(\br)\phi^*_k(\br)$, the Hartree kernel is $f\H(\br,\br')=1/|\br-\br'|$, and $f\xc$ is the XC kernel, here assumed to be frequency independent, which is true in the adiabatic approximation where $f\xc(\br,\br') = \delta(\br-\br')\delta^2 E\xc[\rho]/\delta\rho^2(\br)$. 

The absorption spectrum is proportional to the dipole strength function (DSF) which is written as: 
\begin{equation}
S(\omega) = \sum_q f_q \delta(\omega-\Omega_q) \rightarrow \sum_q f_q \mathcal{L}(\omega,\Omega_q,\Gamma) 
\label{e:dsf}
\end{equation}
where $f_q$ is the oscillator strength, which may be calculated\cite{C96} from the eigenvectors of Eq. (\ref{e:casida}). For comparison to experimental absorption spectra, the DSF is typically broadened using a Lorentzian:
\begin{equation}
\mathcal{L}(\omega,\Omega_q,\Gamma) = \frac{1}{\pi} \frac{ \frac{\Gamma}{2}}{(\omega-\Omega_q)^2+(\frac{\Gamma}{2})^2}
\end{equation}
where $\Gamma=150$ meV was chosen for this work. See Refs.\cite{runge1984,C96,EBF07,U14} for more details on TDDFT.

\subsubsection*{Calculation details}

The DFT and TDDFT calculations for each molecule in the dataset were performed using the Octopus\cite{TOAA20} real-space electronic-structure code. A grid spacing of $0.15$ \AA \ was used in all cases, while the box size was determined by adding a radius of $6.5$ \AA \ around each molecule. The convergence with respect to grid spacing was verified for a number of molecules in the dataset. Norm-conserving, scalar-relativistic, Kleinman-Bylander pseudopotentials were used. The valence density was output in the CUBE file format. In this case, the core density was unavailable to be added, however for visible-UV spectra the valence density should be sufficient (if predicting X-ray instead of vis/UV absorption spectra, adding the density from the core orbitals would likely be necessary). Before the Casida calculation, 200 unoccupied KS orbitals were converged with a tolerance of $1.0\times10^{-7}$ and only KS orbital pairs with energy differences less than $0.75$ au $= 20.4$ eV were included in the Casida diagonalization. 

For this work the local density approximation (LDA) was used for the XC energy functional for the DFT and TDDFT calculations (in the adiabatic approximation). Using the methods detailed in this paper, higher-level XC approximations could be used to construct more accurate ML models for absorption spectra, however as our primary aim is to investigate the density versus geometry input dependence of these ML models, LDA is an appropriate choice. Note that we do not relax the geometry for our simulations and use the geometry given in the QM7 dataset, which was determined via an optimization at the PBE0 level of DFT, which should be closer to the experimental structure. 

Using the frequencies and oscillator strengths from the Casida calculation, the DSF is constructed using Eq. (\ref{e:dsf}) producing a spectrum on a uniform grid containing 2000 frequency points. The machine-learning analysis presented in this work was restricted to the first \(L=900\) bins, corresponding to \(0\leq\omega\leq0.45\) a.u., or approximately \(0\)--\(12.2\) eV. Restricting the target to this interval focuses the model comparison on the spectral window considered relevant to the present molecular and experiment-facing applications while avoiding training the model on higher-energy regions that were outside the scope of the analysis.

We note that, within adiabatic LDA, the Kohn–Sham ionization threshold (-$\varepsilon_{HOMO}$) of these molecules lies at roughly 6-8 eV, so part of the 0-12.2 eV window falls above it. In this region the computed transitions correspond to discretized continuum states whose positions depend on the simulation-box size. The reference spectra are therefore not intended as converged predictions of the experimental vacuum-UV absorption. However, since all spectra were computed with the same protocol, they provide a consistent density-to-spectrum mapping for the purpose of comparing input representations.

\subsection*{Molecular representations and preprocessing}
\label{sec:representations}

\begin{figure}[t]
    \centering
    \includegraphics[width=0.9\linewidth]{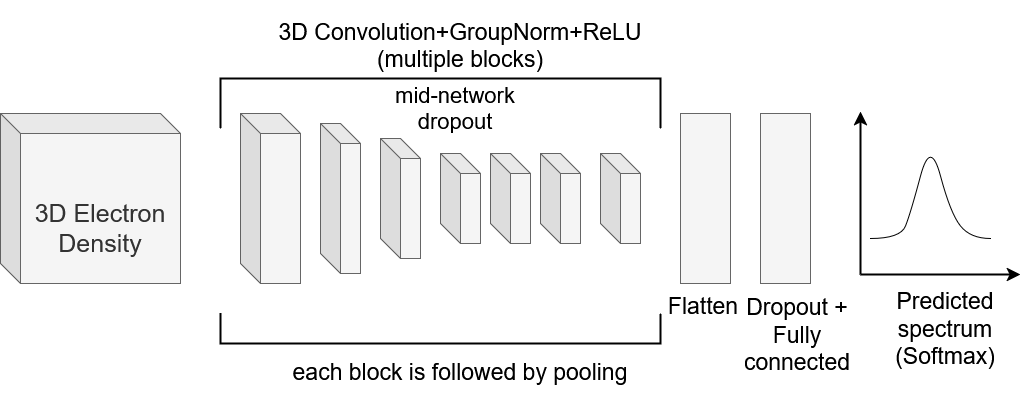}
    \caption{Schematic of the 3D-CNN model mapping an electron-density cube to a normalized spectrum over $L$ frequency bins via a Softmax output.}
    \label{fig:arch_cnn}
\end{figure}

\begin{figure}[t]
    \centering
    \includegraphics[width=0.9\linewidth]{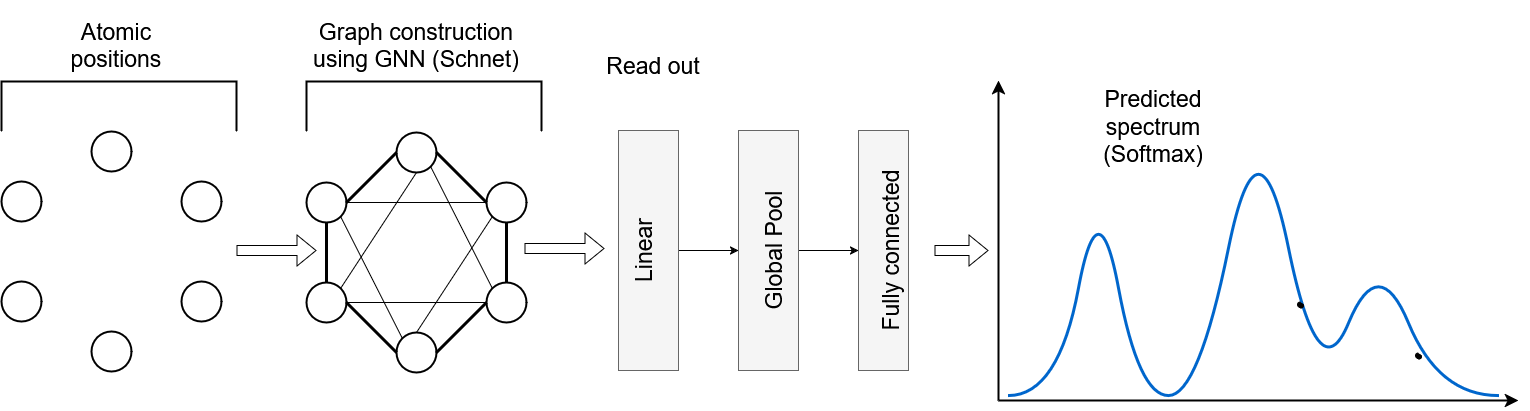}
    \caption{Schematic of the geometry-based GNN model (SchNet / DimeNet++ / MACE) to predict a normalized spectrum.}
    \label{fig:arch_gnn_fusion}
\end{figure}

\subsubsection*{Atomic geometry}
\label{sec:geometry_representation_method}

For the geometry-based models, each molecule was represented by its atomic numbers and three-dimensional nuclear coordinates, as defined in
Eq.~\eqref{eq:geometry_representation}. These inputs require no electronic information beyond the molecular structure and therefore represent the lowest-cost input considered in this study.

Graphs were constructed dynamically from interatomic distances within a model-dependent cutoff radius. Hydrogen atoms were retained explicitly. The cutoff radius, maximum number of neighbours, number of radial basis functions, and other architecture-specific graph settings were treated as model hyperparameters and are reported with the corresponding model configurations.

\subsubsection*{Volumetric electron density}
\label{sec:density_representation}

For the density-based model, the ground-state electron density was represented
on a common Cartesian grid. Each molecular density was stored as a
single-channel tensor,
\begin{equation}
    \boldsymbol{\rho}
    \in
    \mathbb{R}^{1\times155\times147\times143}.
    \label{eq:density_tensor_shape}
\end{equation}
The molecular centre of mass was translated to a common origin, after which
the native density grid was embedded in the fixed-size tensor using
zero-padding. This procedure preserves the spatially resolved density while
allowing all molecules to be processed in fixed-size batches.

The grid spacing inherited from the electronic-structure calculations was
kept fixed across the dataset. Density values were stored using a common
numerical precision and were not reduced to atom-centred or handcrafted
descriptors. 

Because a conventional Cartesian 3D CNN is not intrinsically rotationally
invariant, the orientation inherited from the QM7 geometry and used during
the electronic-structure calculation was retained. The density and its
corresponding spectrum therefore always refer to the same molecular geometry
and orientation convention.

\subsection*{Spectral targets and normalisation}
\label{sec:spectral_targets}

Let
\begin{equation}
    \mathbf{s}
    =
    \left(
        s_1,\ldots,s_L
    \right)
    \in
    \mathbb{R}_{\geq0}^{L}
    \label{eq:raw_spectrum_vector}
\end{equation}
denote the broadened spectrum sampled over the \(L=900\) frequency bins used
for machine learning. To separate relative spectral shape from the overall
intensity scale, we define the integrated discrete spectral weight
\begin{equation}
    \alpha
    =
    \sum_{\ell=1}^{L}s_{\ell},
    \label{eq:spectrum_normalisation_constant}
\end{equation}
and the unit-normalised spectral shape
\begin{equation}
    p_{\ell}
    =
    \frac{s_{\ell}}{\alpha},
    \qquad
    \sum_{\ell=1}^{L}p_{\ell}=1.
    \label{eq:normalised_spectrum}
\end{equation}

The principal spectrum-prediction task in this work is prediction of
\(\mathbf{p}\), which characterises the distribution of spectral intensity
across the explored energy window. This formulation allows the model to focus
on peak positions, relative intensities, shoulders, and the redistribution of
spectral weight without allowing the total intensity scale to dominate the
training objective.

In separate experiments, the normalisation constant \(\alpha\) was treated as
a scalar regression target. A prediction in the original intensity scale may
then be reconstructed as
\begin{equation}
    \widehat{\mathbf{s}}
    =
    \widehat{\alpha}\,
    \widehat{\mathbf{p}}.
    \label{eq:reconstructed_spectrum}
\end{equation}
The main model comparisons reported for the spectral task concern the
normalised spectral shape unless explicitly stated otherwise.

\subsection*{Choice of neural architectures}
\label{sec:model_choice}

The neural architectures were selected to represent distinct physical and
practical compromises rather than to constitute an unstructured benchmark of
available molecular models. Geometry-based graph neural networks require only
atomic species and nuclear coordinates. They can therefore be applied directly
to candidate structures and are suitable for rapid screening before an
electronic-structure calculation has been performed.

The density-based CNN requires a ground-state DFT calculation and is therefore
not an end-to-end replacement for electronic-structure theory. Instead, it
addresses the more expensive response stage by learning the mapping from the
ground-state density to the absorption spectrum. This places it between
geometry-only machine learning and explicit LR-TDDFT: its input is more
expensive to generate than a molecular graph, but contains direct electronic
information that the graph models must infer from nuclear geometry.

The graph architectures were chosen to provide a hierarchy of geometric
inductive biases. SchNet provides a distance-based continuous-filter baseline,
DimeNet++ adds explicit angular and directional information, and MACE
introduces symmetry-adapted higher-order equivariant features. Comparing these
models with the density CNN tests whether increasingly expressive
geometry-based representations can recover the information present explicitly
in the electron density.

\subsection*{Neural-network architectures}
\label{sec:architectures}

All spectrum models output \(L=900\) logits followed by a Softmax operation,
ensuring a non-negative unit-normalised prediction,
\begin{equation}
    \widehat{p}_{\ell}
    =
    \frac{\exp(z_{\ell})}
    {\sum_{j=1}^{L}\exp(z_j)}.
    \label{eq:softmax_spectrum}
\end{equation}
The identical output representation was used across model families so that differences in performance could be attributed to the learned representations and architectures rather than to different spectral normalisation procedures.

\subsubsection*{Three-dimensional convolutional neural network}
\label{sec:cnn_architecture}

A 3D CNN is a natural architecture for processing the electron density because the input is defined on a regular Cartesian grid. Spatially shared convolutional filters can identify recurring local density patterns, while successive layers combine these local patterns over progressively larger spatial regions. This allows the model to retain the full volumetric representation without first reducing it to a predetermined set of electronic descriptors.

The density tensor was processed by a sequence of 3D convolutional blocks. Each block consisted of a three-dimensional convolution, group normalisation, a nonlinear activation, and spatial pooling. Dropout was applied within the convolutional stack and before the final prediction layer. The final feature maps were flattened rather than globally averaged so that the prediction head retained the learned spatial arrangement of the density features. The spectrum prediction can be written schematically as
\begin{equation}
    \widehat{\mathbf{p}}
    =
    \operatorname{Softmax}
    \left[
        \mathbf{W}_{\rho}
        \operatorname{Flatten}
        \left(
            \operatorname{CNN}
            (\boldsymbol{\rho})
        \right)
        +
        \mathbf{b}_{\rho}
    \right].
    \label{eq:cnn_spectrum}
\end{equation}
The number of convolutional layers, convolutional kernel size, initial number
of filters, and dropout probability were varied during the hyperparameter
search. The configurations retained for comparison are reported in the
corresponding results table.

\subsubsection*{SchNet}
\label{sec:schnet_architecture}

SchNet represents a molecule through atom-centred latent features updated by
continuous-filter convolutions over interatomic distances
\cite{schutt2017schnet,schutt2018schnet}. It provides a computationally compact
geometry-based baseline that is invariant to global translations and rotations
of the molecular coordinates. Its distance-based construction tests how well
the spectrum can be inferred from radial structural information without
explicit angular features.

Starting from embeddings of the atomic numbers, SchNet updates the atomic
features using neighbours within a cutoff radius. The final atomic
representations \(\{\mathbf{h}_i\}\) are aggregated to form a molecular
embedding, which is mapped to the spectral bins:
\begin{equation}
    \widehat{\mathbf{p}}
    =
    \operatorname{Softmax}
    \left[
        \mathbf{W}_{\mathrm{S}}
        \operatorname{Pool}
        \left(
            \{\mathbf{h}_i\}_{i=1}^{N_{\mathrm{at}}}
        \right)
        +
        \mathbf{b}_{\mathrm{S}}
    \right].
    \label{eq:schnet_spectrum}
\end{equation}

In addition to a conventional graph-level readout, we evaluated an
attention-enhanced multi-pooling variant. This readout concatenates mean, sum,
and maximum pooling with an attention-weighted graph embedding before the
final spectral prediction layer. The purpose of this variant was to test
whether a more expressive global aggregation procedure improves the mapping
from local atom-centred features to a nonlocal molecular response.

\subsubsection*{DimeNet++}
\label{sec:dimenet_architecture}

Distance-only representations do not explicitly distinguish environments with
similar interatomic distances but different bond angles. DimeNet addresses
this limitation through directional message passing using radial and angular
basis functions \cite{gasteiger2020dimenet}. We use the computationally
improved DimeNet++ architecture \cite{gasteiger2020dimenetpp}, which retains
the directional inductive bias while reducing the computational cost.

DimeNet++ was included to test whether explicit angular information improves
spectral prediction relative to the distance-based SchNet representation.
The molecular embedding produced by the directional interaction blocks was
mapped to \(L\) spectral logits and normalised using the same Softmax head as
the other spectrum models:
\begin{equation}
    \widehat{\mathbf{p}}
    =
    \operatorname{Softmax}
    \left[
        \mathbf{W}_{\mathrm{D}}
        \mathbf{h}_{\mathrm{mol}}^{\mathrm{D}}
        +
        \mathbf{b}_{\mathrm{D}}
    \right].
    \label{eq:dimenet_spectrum}
\end{equation}

\subsubsection*{MACE}
\label{sec:mace_architecture}

MACE combines equivariant message passing with the atomic cluster expansion
and can propagate scalar, vector, and higher-order tensor features
\cite{batatia2022mace,drautz2019ace}. The architecture therefore provides a
strong test of whether symmetry-aware and higher-body geometric information
can reduce the difference between geometry-based and density-based spectrum
prediction.

The hidden features are described using the notation \(n\times\ell_p\), where
\(n\) is the number of channels, \(\ell\) denotes the angular order, and \(p\)
denotes parity under spatial inversion. Scalar features correspond to
\(\ell=0\), vector-like features to \(\ell=1\), and higher values of \(\ell\)
to higher-order equivariant tensors. Tensor products and spherical harmonics
were implemented using the e3nn framework \cite{geiger2022e3nn}.

For example,
\begin{equation}
    128\times0e + 128\times1o
\end{equation}
denotes 128 even scalar channels and 128 odd vector channels. The architecture
search varied the number of interaction blocks, channel multiplicities,
dropout, and the highest angular order represented in the hidden features.

Only rotationally invariant molecular features were passed to the scalar
spectrum head. The resulting prediction is
\begin{equation}
    \widehat{\mathbf{p}}
    =
    \operatorname{Softmax}
    \left[
        \mathbf{W}_{\mathrm{M}}
        \mathbf{h}_{\mathrm{mol}}^{\mathrm{M}}
        +
        \mathbf{b}_{\mathrm{M}}
    \right],
    \label{eq:mace_spectrum}
\end{equation}
where \(\mathbf{h}_{\mathrm{mol}}^{\mathrm{M}}\) is obtained through
permutation-invariant aggregation of the molecular features.

\subsection*{Loss functions}
\label{sec:losses}

Because the spectral target is a unit-normalised non-negative vector, we
evaluated both distributional and pointwise learning objectives. Let
\(\mathbf{p}\) denote the reference spectral distribution and
\(\widehat{\mathbf{p}}\) the prediction.

(a) The Kullback--Leibler divergence is
\begin{equation}
    D_{\mathrm{KL}}
    \left(
        \mathbf{p}
        \parallel
        \widehat{\mathbf{p}}
    \right)
    =
    \sum_{\ell=1}^{L}
    p_{\ell}
    \log
    \left(
        \frac{p_{\ell}}
        {\widehat{p}_{\ell}}
    \right).
    \label{eq:kl_loss}
\end{equation}
A small positive constant was applied inside logarithms to avoid numerical
instability for zero-valued bins.

(b) The Jensen--Shannon divergence is the symmetric, bounded quantity
\cite{lin1991jsd}
\begin{align}
    \mathbf{m}
    &=
    \frac{1}{2}
    \left(
        \mathbf{p}
        +
        \widehat{\mathbf{p}}
    \right),
    \\
    D_{\mathrm{JS}}
    \left(
        \mathbf{p}
        \parallel
        \widehat{\mathbf{p}}
    \right)
    &=
    \frac{1}{2}
    D_{\mathrm{KL}}
    \left(
        \mathbf{p}
        \parallel
        \mathbf{m}
    \right)
    +
    \frac{1}{2}
    D_{\mathrm{KL}}
    \left(
        \widehat{\mathbf{p}}
        \parallel
        \mathbf{m}
    \right).
    \label{eq:js_loss}
\end{align}

(c) The pointwise mean absolute and mean squared errors are
\begin{align}
    \mathcal{L}_{\mathrm{MAE}}
    &=
    \frac{1}{L}
    \sum_{\ell=1}^{L}
    \left|
        p_{\ell}
        -
        \widehat{p}_{\ell}
    \right|,
    \label{eq:mae_loss}
    \\
    \mathcal{L}_{\mathrm{MSE}}
    &=
    \frac{1}{L}
    \sum_{\ell=1}^{L}
    \left(
        p_{\ell}
        -
        \widehat{p}_{\ell}
    \right)^2.
    \label{eq:mse_loss}
\end{align}

(d) Cosine loss measures the agreement in the global direction of the two spectral
vectors,
\begin{equation}
    \mathcal{L}_{\mathrm{cos}}
    =
    1
    -
    \frac{
        \mathbf{p}\cdot\widehat{\mathbf{p}}
    }{
        \|\mathbf{p}\|_2
        \|\widehat{\mathbf{p}}\|_2
    },
    \label{eq:cosine_loss}
\end{equation}
while Pearson-correlation loss is defined as
\begin{equation}
    \mathcal{L}_{\mathrm{corr}}
    =
    1-r
    \left(
        \mathbf{p},
        \widehat{\mathbf{p}}
    \right),
    \label{eq:pearson_loss}
\end{equation}
where \(r\) is the Pearson correlation coefficient evaluated across frequency
bins.

(e) To account explicitly for displacement of spectral weight along the frequency
axis, we also used the one-dimensional Wasserstein-1 distance. Defining the
discrete cumulative distributions
\begin{equation}
    F_{\mathbf{p}}(k)
    =
    \sum_{\ell=1}^{k}p_{\ell},
    \qquad
    F_{\widehat{\mathbf{p}}}(k)
    =
    \sum_{\ell=1}^{k}\widehat{p}_{\ell},
    \label{eq:cumulative_spectra}
\end{equation}
the discrete Wasserstein loss is
\begin{equation}
    \mathcal{L}_{W_1}
    =
    \sum_{k=1}^{L}
    \left|
        F_{\mathbf{p}}(k)
        -
        F_{\widehat{\mathbf{p}}}(k)
    \right|.
    \label{eq:wasserstein_loss}
\end{equation}
Equation~\eqref{eq:wasserstein_loss} is expressed in frequency-bin units.
Multiplication by the frequency spacing \(\Delta\omega\) converts it to the
corresponding frequency units. This distinction is retained
when interpreting the reported Wasserstein values.

For scalar targets, including the gap quantities and spectral normalisation
constant, standard MAE or MSE regression objectives were used. All losses were
averaged over the molecules in each minibatch.

\subsection*{Training protocol and hyperparameter selection}
\label{sec:training}

The molecular dataset was shuffled at the molecule level using a fixed random
seed and divided into \(85\%\) training data and \(15\%\) validation data. The
same indices were used for the density CNN and all geometry-based models.
Consequently, no molecular structure appeared in both subsets, and all
architecture comparisons were evaluated on an identical validation set.

No additional test partition was introduced. The validation subset was used
for hyperparameter selection and for the comparative results reported in this
work. Accordingly, all reported quantities are described explicitly as
validation performance rather than independent test-set estimates.

The models were implemented using PyTorch, with graph construction and graph
batching performed using PyTorch Geometric
\cite{fey2019pytorchgeometric}. Optimisation used AdamW. The model families
required different batch sizes because of their substantially different memory
requirements, particularly for the volumetric density tensors. Within each
model family, the optimisation protocol and validation procedure were kept
fixed across the corresponding hyperparameter trials.

The CNN search varied the number of convolutional layers, convolution kernel
size, initial number of filters, and dropout probability. The SchNet search
varied the hidden dimension, number of interaction blocks, number of Gaussian
basis functions, cutoff radius, learning rate, and training objective. The
DimeNet++ search varied the hidden dimension, number of interaction blocks,
radial basis size, cutoff radius, maximum number of neighbours, learning rate,
and loss function. The MACE search examined the hidden irreducible
representations, interaction depth, angular order, dropout, learning rate, and
training objective.

\subsection*{Evaluation metrics}
\label{sec:evaluation}

Spectrum prediction was evaluated using complementary metrics that quantify
different forms of spectral agreement. For each molecule, Pearson correlation
was calculated across the \(L=900\) frequency bins,
\begin{equation}
    r
    =
    \frac{
        \sum_{\ell}
        \left(
            p_{\ell}-\overline{p}
        \right)
        \left(
            \widehat{p}_{\ell}
            -
            \overline{\widehat{p}}
        \right)
    }{
        \sqrt{
            \sum_{\ell}
            \left(
                p_{\ell}-\overline{p}
            \right)^2
        }
        \sqrt{
            \sum_{\ell}
            \left(
                \widehat{p}_{\ell}
                -
                \overline{\widehat{p}}
            \right)^2
        }
    }.
    \label{eq:pearson_metric}
\end{equation}
Pearson correlation measures whether the prediction reproduces the overall line shape and relative variation of intensity across the spectral range. The model-level validation correlation was obtained by averaging the per-molecule correlations over the validation set. Median correlations were used for the
chemical-category analysis to reduce sensitivity to individual outliers.

The spectral MAE was calculated over frequency bins and averaged over the validation molecules. It measures the pointwise discrepancy in the unit-normalised spectral intensity and is sensitive to incorrect peak heights, residual background intensity, and errors in weak spectral features. KL and JS divergences quantify distributional agreement, while Wasserstein distance measures the displacement or redistribution of spectral intensity along the frequency axis.

Using several metrics is necessary because no single quantity captures every spectral failure mode. A prediction may reproduce the broad line shape and therefore attain a high correlation while retaining local amplitude errors. Conversely, a low average pointwise error may be obtained by an excessively smooth prediction that does not reproduce weaker peaks or shoulders.

Finally, validation performance was stratified by chemical categories,
including bond types, unsaturation, aromaticity, heteroatom content, rings,
and functional groups. For each category, molecules in which the feature was
present were compared with those in which it was absent. These comparisons
were used to determine whether the reported aggregate accuracy was maintained
across distinct chemical environments and to identify under-represented or
electronically challenging molecular classes.

\section*{Results}

\subsection*{Prediction of normalized absorption spectra}

We first train the models and evaluate the prediction of absorption spectra on the unseen validation set. The target in this task is the normalized spectral distribution over $L=900$ frequency bins. Since the model output is a Softmax-normalized spectrum, performance is assessed using complementary metrics: Pearson correlation for spectral shape, MAE for pointwise intensity error, KL/JS divergence for distributional agreement, and Wasserstein distance for redistribution of spectral mass along the frequency axis.

We begin with the graph-based baselines. Table~\ref{tab:schnet_dimenet_sweeps} compares SchNet and DimeNet++ across loss functions and hyperparameter settings. Both architectures achieve strong validation performance when trained with KL divergence, JS divergence, L1 loss, or cosine loss. The best SchNet model reaches a validation correlation of $0.9795$ with a validation MAE of $1.36\times10^{-4}$. DimeNet++ gives comparable performance, with the best KL-trained configurations reaching validation correlations of approximately $0.9794$ and MAE values close to $1.36\times10^{-4}$. These results show that both distance-based and directional message-passing models can learn the global spectral distribution from atom types and nuclear coordinates.

\begin{table*}[t]
\caption{Validation performance for graph-based spectrum predictors (SchNet and DimeNet++). ``Int.'' denotes SchNet interaction blocks; DimeNet++ columns correspond to \texttt{num\_blocks} (Blocks), \texttt{num\_radial} (Radial), and maximum neighbors (MaxNN). Because loss functions have different numeric scales (e.g., Wasserstein vs KL/JS), correlation and MAE provide the most comparable cross-run summary}
\label{tab:schnet_dimenet_sweeps}
\centering
\subfloat[SchNet sweeps \cite{schutt2017schnet}. Best corr and best MAE are bolded.]{
\begin{tabular}{r c c c c c l c c c}
\toprule
Run & Hidden & Int. & Gauss. & Cutoff & LR & Loss & Val loss & Val corr & Val MAE \\
\midrule
1 & 64  & 8 & 25 & 5.0 & 8.53e-04 & wasserstein & 4.012  & 0.8216 & 4.47e-04 \\
2 & 64  & 6 & 50 & 5.0 & 1.47e-04 & kl          & 0.0166 & \textbf{0.9795} & 1.36e-04 \\
3 & 128 & 4 & 50 & 3.0 & 3.25e-04 & l1          & 1.37e-04 & 0.9788 & 1.37e-04 \\
4 & 64  & 4 & 50 & 3.0 & 2.88e-04 & cosine      & 0.0117 & 0.9784 & 1.43e-04 \\
5 & 64  & 3 & 50 & 5.0 & 4.29e-05 & kl          & 0.0206 & 0.9754 & 1.50e-04 \\
6 & 128 & 8 & 50 & 6.0 & 3.32e-05 & l1          & 1.40e-04 & 0.9773 & 1.40e-04 \\
7 & 32  & 3 & 75 & 5.0 & 2.63e-04 & js          & 0.00447 & 0.9776 & 1.42e-04 \\
8 & 128 & 8 & 50 & 3.0 & 2.89e-04 & js          & 0.00406 & 0.9794 & \textbf{1.36e-04} \\
9 & 32  & 4 & 25 & 3.0 & 4.04e-04 & wasserstein & 4.501  & 0.6696 & 7.35e-04 \\
10& 32  & 3 & 50 & 6.0 & 9.91e-04 & wasserstein & 4.424  & 0.7344 & 6.15e-04 \\
\bottomrule
\end{tabular}
}
\qquad
\subfloat[DimeNet++ sweeps \cite{gasteiger2020dimenetpp}. Best corr and best MAE are bolded.]{
\begin{tabular}{r c c c c c c l c c c}
\toprule
Run & Hidden & Blocks & Radial & Cutoff & MaxNN & LR & Loss & Val loss & Val corr & Val MAE \\
\midrule
1 & 64  & 4 & 24 & 5.0 & 64 & 1.78e-05 & cosine & 0.0133 & 0.9753 & 1.59e-04 \\
2 & 256 & 3 & 24 & 6.0 & 64 & 3.76e-05 & js     & 0.00424 & 0.9786 & 1.39e-04 \\
3 & 256 & 6 & 12 & 6.0 & 16 & 3.81e-05 & kl     & 0.0168 & \textbf{0.9794} & 1.37e-04 \\
4 & 64  & 2 & 24 & 6.0 & 32 & 8.98e-05 & js     & 0.00442 & 0.9779 & 1.42e-04 \\
5 & 256 & 3 & 6  & 3.0 & 16 & 6.26e-04 & l1     & 1.36e-04 & 0.9790 & 1.36e-04 \\
6 & 256 & 6 & 24 & 6.0 & 32 & 1.11e-04 & kl     & 0.0163 & 0.9794 & \textbf{1.36e-04} \\
7 & 64  & 6 & 24 & 5.0 & 32 & 9.67e-05 & cosine & 0.0118 & 0.9782 & 1.50e-04 \\
8 & 64  & 4 & 24 & 3.0 & 32 & 1.49e-04 & cosine & 0.0117 & 0.9784 & 1.48e-04 \\
9 & 64  & 3 & 6  & 6.0 & 32 & 1.72e-05 & kl     & 0.0179 & 0.9782 & 1.41e-04 \\
10& 128 & 3 & 12 & 5.0 & 16 & 1.59e-05 & kl     & 0.0182 & 0.9780 & 1.40e-04 \\
\bottomrule
\end{tabular}
} 
\end{table*}

\begin{table*}[ht]
\centering
\setlength{\tabcolsep}{2pt}
\footnotesize
\caption{MACE sweep on spectra prediction ($L=900$ bins). ``Int.'' denotes the number of interaction (message-passing) blocks}
\begin{tabular}{l l c c c c c c}
\toprule
Exp. & Hidden irreps & Int. & $\ell_{\max}$ & LR & Dropout & Val loss & Val corr \\
\midrule
exp01 & 128x0e + 128x1o & 2 & 3 & $8\times10^{-5}$ & 0.10 & \textbf{0.0171086} & 0.979423 \\
exp02 & 128x0e + 128x1o & 3 & 3 & $8\times10^{-5}$ & 0.10 & 0.0171974 & 0.979223 \\
exp03 & 128x0e + 128x1o & 4 & 3 & $8\times10^{-5}$ & 0.10 & 0.0174657 & 0.978854 \\
exp04 & 192x0e + 192x1o & 2 & 3 & $8\times10^{-5}$ & 0.10 & 0.0171763 & \textbf{0.979475} \\
exp05 & 96x0e + 96x1o + 96x2e & 2 & 3 & $8\times10^{-5}$ & 0.10 & 0.0175385 & 0.979077 \\
exp06 & 128x0e (scalar control) & 2 & 3 & $8\times10^{-5}$ & 0.10 & 0.0183623 & 0.977993 \\
exp07 & 128x0e + 128x1o & 2 & 3 & $5\times10^{-5}$ & 0.10 & 0.0179472 & 0.978512 \\
exp08 & 128x0e + 128x1o & 2 & 3 & $8\times10^{-5}$ & 0.20 & 0.0171181 & 0.979426 \\
\bottomrule
\end{tabular}
\label{tab:mace_sweep}
\end{table*}

\begin{table*}[htb]
\centering
\setlength{\tabcolsep}{2pt}
\footnotesize
\caption{Loss-function comparison for the selected MACE spectrum model with hidden irreps $128\times0e + 128\times1o$, two interaction blocks, $\ell_{\max}=3$, $r_{\max}=5.0$, eight radial Bessel functions, learning rate $8\times10^{-5}$, and dropout 0.10. For each row, the model is trained using the loss shown in the first column, while all other quantities are evaluated as validation metrics. JS and KL divergence give the best balanced performance across correlation, pointwise error, and distributional metrics}
\begin{tabular}{l c c c c c c c}
\toprule
Training loss  & Val MAE & Val MSE & Val corr & Val KL & Val JS & Val $W_1$ \\
\midrule
Wasserstein  & $5.12\times10^{-4}$ & $6.81\times10^{-7}$ & 0.826412 & 0.723892 & 0.069020 & \textbf{4.307083} \\
Pearson corr. & $8.45\times10^{-4}$ & $9.98\times10^{-7}$ & 0.977388 & 0.435214 & 0.125681 & 205.415649 \\
JS divergence & $\textbf{1.38}\times10^{-4}$ & $\textbf{6.17}\times10^{-8}$ & \textbf{0.979338} & 0.017322 & \textbf{0.004197} & 4.735388 \\
KL divergence  & $1.39\times10^{-4}$ & $6.20\times10^{-8}$ & 0.979314 & \textbf{0.017253} & 0.004210 & 4.724957 \\
MAE  & $1.43\times10^{-4}$ & $6.89\times10^{-8}$ & 0.976748 & 0.020495 & 0.004796 & 5.737404 \\
MSE & $2.06\times10^{-4}$ & $1.28\times10^{-7}$ & 0.959622 & 0.045174 & 0.010397 & 16.948169 \\
\bottomrule
\end{tabular}
\label{tab:mace_loss_sweep}
\end{table*}

\begin{table}[hbt]
\centering
\setlength{\tabcolsep}{2pt}
\footnotesize
\caption{Top-5 3D-CNN density-to-spectrum models ranked by validation correlation. The validation loss corresponds to the mean KL divergence reported on the validation set. Loss and MAE values are shown in scientific notation and rounded to three significant figures; correlations are rounded to four decimal places}
\label{tab:cnn_top5}
\begin{tabular}{r r r r r r r r}
\toprule
Rank & Layers & Kernel & BaseF & Dropout & Val loss & Val corr & Val MAE \\
\midrule
1 & 7  & 15 & 8 & 0.20 & 6.66e-06 & 0.9926 & 7.14e-05 \\
2 & 7  & 7  & 6 & 0.10 & 7.18e-06 & 0.9922 & 7.54e-05 \\
3 & 8  & 5  & 4 & 0.10 & 1.84e-05 & 0.9794 & 1.44e-04 \\
4 & 4  & 7  & 6 & 0.30 & 2.59e-05 & 0.9757 & 1.57e-04 \\
5 & 10 & 9  & 8 & 0.30 & 2.98e-05 & 0.9682 & 1.79e-04 \\
\bottomrule
\end{tabular}
\end{table}

The loss-function dependence in Table~\ref{tab:schnet_dimenet_sweeps} is also important. KL, JS, L1, and cosine objectives all lead to high correlations, whereas the Wasserstein-trained SchNet runs give substantially lower correlations. This indicates that directly optimizing the cumulative-distribution error does not necessarily produce the best pointwise or correlation-based spectral agreement. Overall, the SchNet and DimeNet++ sweeps suggest that, within the graph-based representation, performance is affected at least as strongly by the training objective as by the architectural distinction between the two best graph models.

We next consider the equivariant MACE model. The architecture sweep in Table~\ref{tab:mace_sweep} shows that MACE reaches the same overall performance regime as SchNet and DimeNet++. The best validation correlation is obtained by the model with hidden irreps \texttt{192x0e + 192x1o} and two interaction blocks, giving a validation correlation of $0.979475$. The lowest validation loss is obtained by the \texttt{128x0e + 128x1o} model with two interaction blocks, which reaches a validation loss of $0.01710$ and a correlation of $0.979$. Increasing the number of interaction blocks from two to three or four does not improve the result, suggesting that the relevant structural information for this dataset is already captured by relatively shallow message passing. The scalar-only control model gives a lower correlation of $0.978$, indicating that vector equivariant features provide a small but measurable benefit.

A direct comparison of training objectives for MACE is given in Table~\ref{tab:mace_loss_sweep}. For the selected MACE architecture, JS and KL divergence give the best balanced performance. JS divergence gives the best validation correlation, $0.979338$, the lowest MAE, $1.38\times10^{-4}$, and the lowest JS divergence. KL divergence gives a nearly identical correlation of $0.979314$ and the best validation KL value. In contrast, MSE gives a noticeably lower correlation of $0.959622$, while Wasserstein training gives the lowest $W_1$ value but a much lower correlation of $0.826412$. These results support the use of distributional objectives for Softmax-normalized spectra.

As the networks have so far shown the best performance with KL divergence loss, the density-based CNNs were trained using this loss function, and they outperformed the graph-based neural networks on the full-spectrum task. As shown in Table~\ref{tab:cnn_top5}, the best CNN configuration reaches a validation correlation of $0.9926$, a validation MAE of $7.14\times10^{-5}$, and a mean validation KL divergence of $6.66\times10^{-6}$. This is a substantial improvement over the best graph-based spectrum predictors. Although the absolute increase in correlation relative to the best graph models is around $0.013$, correlation is bounded above by one, so the more informative quantity is the remaining decorrelation, $1-r$. On this scale, the CNN reduces the residual decorrelation from approximately $2.0\times10^{-2}$ for the best graph models to $7.4\times10^{-3}$, removing nearly two-thirds of the remaining gap to perfect spectral-shape agreement. The improvement is also clearly reflected in the other metrics as CNN reduces the MAE by approximately a factor of two relative to the best graph models, and the reported KL divergence is orders of magnitude smaller than the KL values obtained by the graph-based spectra models.

\begin{figure}[t]
    \centering
    \includegraphics[width=0.9\linewidth]{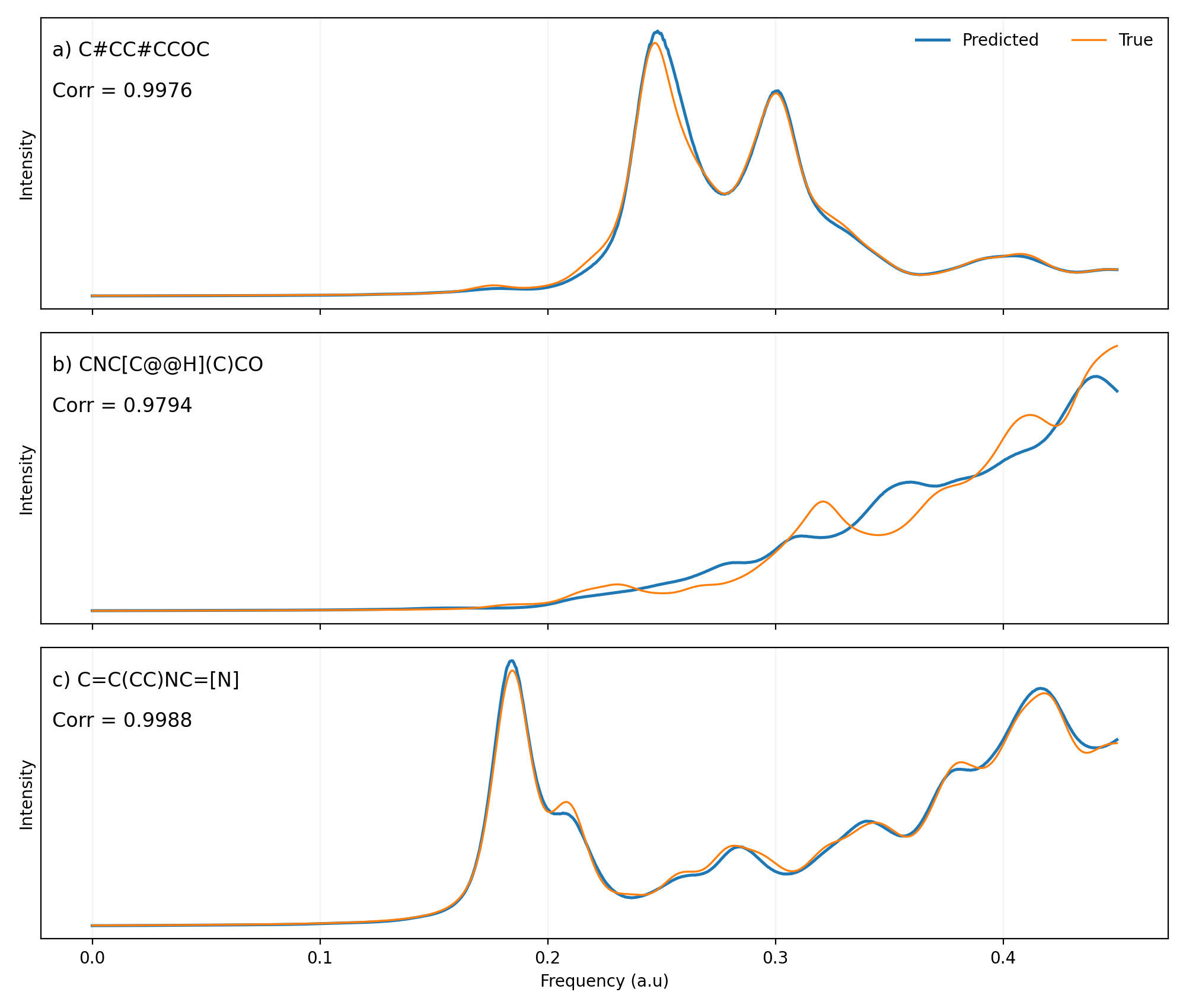}
    \caption{Representative examples of predicted absorption spectra compared with the corresponding ground-truth spectra. Each panel shows one validation molecule using Simplified Molecular Input Line Entry System (SMILES) notation\cite{Weininger1988}: (a) C\#CC\#CCOC, (b) CNC[C@@H](C)CO, and (c) C=C(CC)NC=[N]. The horizontal axis is frequency in atomic units (a.u.) and the vertical axis is the spectral intensity. For each molecule, the model prediction (blue) is overlaid with the reference spectrum (orange), enabling a direct visual comparison of peak positions, relative amplitudes, and overall spectral shape across the full frequency range. The Pearson correlation coefficient between the predicted and true spectra is reported within each panel (Corr), computed over all frequency bins shown. These examples illustrate that the model captures the dominant spectral features and broad trends, while deviations remain in low amplitude peaks and the finer structure.} %
    \label{fig:spectra_val_triptych_0488_0277_0317}
\end{figure}

This result is also physically meaningful. The graph models use atom types and nuclear coordinates, and must therefore infer the relevant electronic structure indirectly from the molecular geometry. In contrast, the 3D-CNN operates directly on the electron-density, which is closer to the physical object that determines optical response. The density representation therefore provides direct spatial information about bonding, charge distribution, local electronic environments, and electronic delocalization. The CNN can exploit these local and extended density patterns through convolutional filters, which explains why the density-based representation is particularly effective for predicting the full spectral distribution.

The architecture schematics in Figs.~\ref{fig:arch_cnn} and~\ref{fig:arch_gnn_fusion} illustrate this representation-level distinction. The graph models learn spectra from a compact atom-centred representation, whereas the CNN receives a volumetric electronic field. The results in Tables~\ref{tab:schnet_dimenet_sweeps}--\ref{tab:cnn_top5} show that both representations are highly informative, but the electron-density representation gives the most accurate normalized absorption spectra.

\begin{figure}[t]
    \centering
    \includegraphics[width=0.9\linewidth]{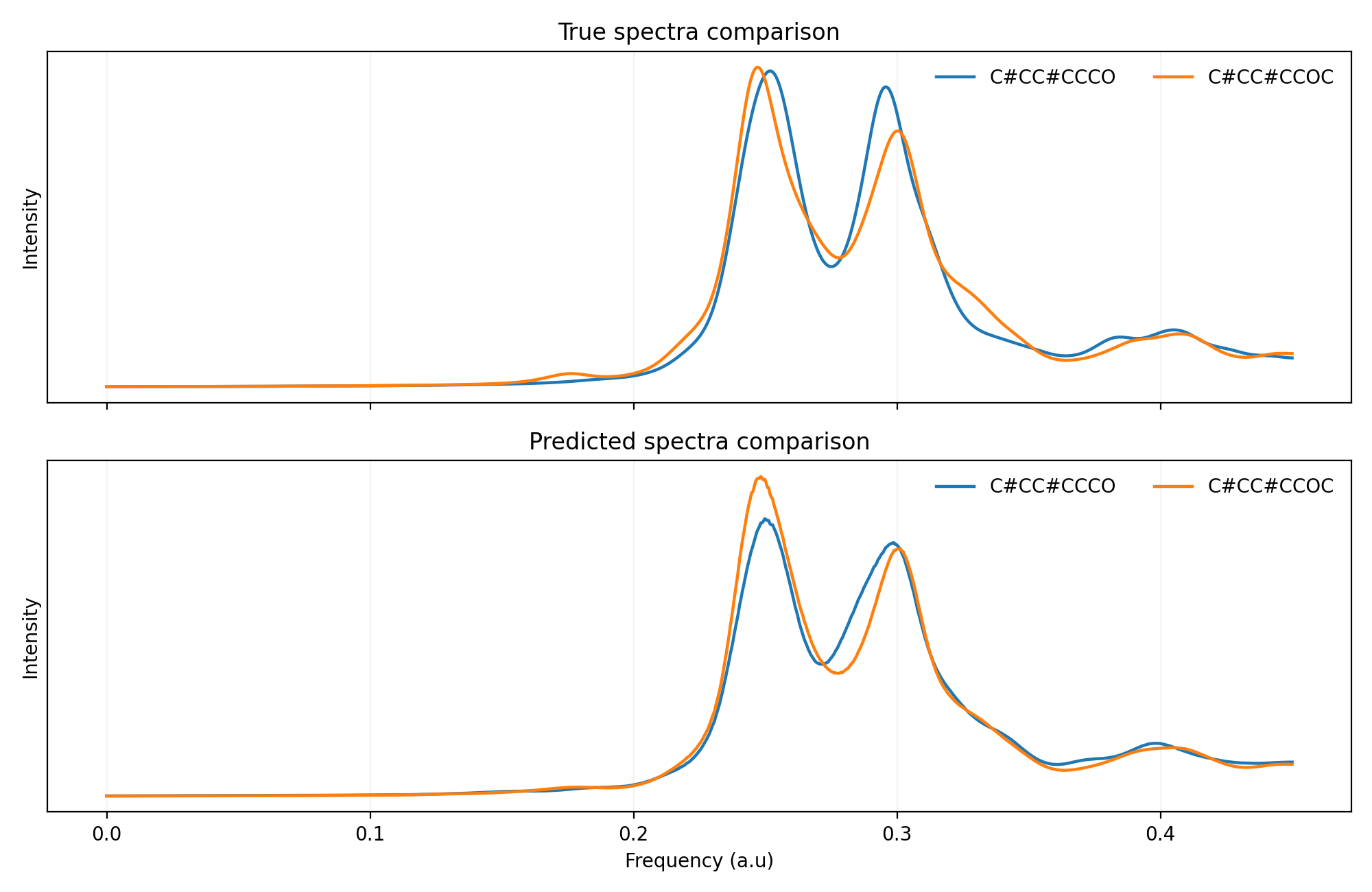}
    \caption{Comparison of two molecules with similar spectral structure and similar molecules, shown separately for the reference spectra and the model outputs. The top panel overlays the ground-truth absorption spectra for the two validation molecules C\#CC\#CCCO and C\#CC\#CCOC as a function of frequency (a.u.), highlighting their closely related peak positions and relative intensities. The bottom panel shows the corresponding model-predicted spectra for the same two molecules, plotted over the same frequency range. The predictions reproduce the overall bimodal profile present in both molecules and capture subtle inter-molecular differences. For instance, the peak and region near 0.3$\sim$a.u. has a larger amplitude for C\#CC\#CCOC in the ground truth, and the same trend is reflected in the model output.}
    \label{fig:compare_true_pred_0221_0488}
\end{figure}

Representative spectra are shown in Fig.~\ref{fig:spectra_val_triptych_0488_0277_0317}. The predicted spectra from the CNN model reproduce the dominant absorption features, broad spectral shapes, and relative intensity trends across molecules with different spectral morphologies. The remaining discrepancies are mainly associated with weak peaks, shoulder-like features, and subtle local structure. The comparison between chemically similar molecules in Fig.~\ref{fig:compare_true_pred_0221_0488} further shows that the model is not simply learning an averaged spectrum. The two molecules have closely related spectra, but the model preserves the relative change in peak amplitude near $0.3$~a.u., indicating sensitivity to small molecular modifications.

This comparison is also relevant from an analytical-chemistry perspective. In gas-phase molecular spectroscopy, candidate assignment often requires distinguishing between structurally related species, fragments, conformers, or tautomers with similar spectral features. The fact that the model preserves subtle relative-intensity differences between two closely related molecules suggests that fast ML-generated spectra could be useful as reference spectra for molecular identification within the chemical space and frequency window studied here. Although mixture deconvolution, concentration effects, and time-dependent tautomer populations are not explicitly modeled in the present work, rapid generation of candidate spectra is a necessary component of such workflows and could support subsequent fitting or assignment procedures.

Taken together, these results show that graph-based models provide strong baselines for spectrum prediction, but the density-based CNN is the best model for the absorption-spectrum task. This supports the central representation-level conclusion that electron density is a more direct and physically appropriate input for predicting molecular absorption spectra than atom-centred geometry alone.

\squeezetable
\begin{table}[t]
    \caption{Computational characteristics of the selected best-performing spectrum-prediction models. Training times correspond to the estimated wall-clock duration of the complete training runs and exclude the wider hyperparameter searches. All models were trained using a single NVIDIA A100 GPU with 40~GB of memory}
    \label{tab:model_compute}
    \begin{tabular}{lrrl}
        \hline
        Model & Parameters & Training time (h) & Hardware \\
        \hline
        SchNet         & 228,534     & 2.30  & NVIDIA A100 40 GB \\
        DimeNet++      & 6,545,808   & 3.80  & NVIDIA A100 40 GB \\
        MACE           & 2,814,211   & 3.22  & NVIDIA A100 40 GB \\
        Density 3D-CNN & 401,295,280 & 91.07 & NVIDIA A100 40 GB \\
        \hline
    \end{tabular}
\end{table}

\begin{table}[t]
\setlength{\tabcolsep}{2pt}
\footnotesize
    \centering
    \caption{
        Batch-size-one inference performance of the selected
        spectrum-prediction models on an NVIDIA A100 GPU.
        Latency denotes the average inference time per molecule, while
        throughput gives the corresponding number of molecular predictions
        per second. GPU memory denotes the peak CUDA memory allocated during
        the post-hoc inference benchmark
    }
    \label{tab:inference_benchmark}
    \begin{tabular}{lrrr}
        \hline
        Model
        & Latency
        & Throughput
        & GPU memory \\
        & (ms molecule$^{-1}$)
        & (molecules s$^{-1}$)
        & (MB) \\
        \hline
        SchNet         & 4.69  & 213.15 & $\sim$10 \\
        MACE           & 12.44 & 80.39  & 106       \\
        Density 3D-CNN & 43.95 & 22.75  & 2397      \\
        \hline
    \end{tabular}
\end{table}

\subsection*{Prediction of the spectral normalisation constant}
\label{sec:normalisation_results}

\begin{figure}[t]
    \centering
    \includegraphics[width=0.9\linewidth]{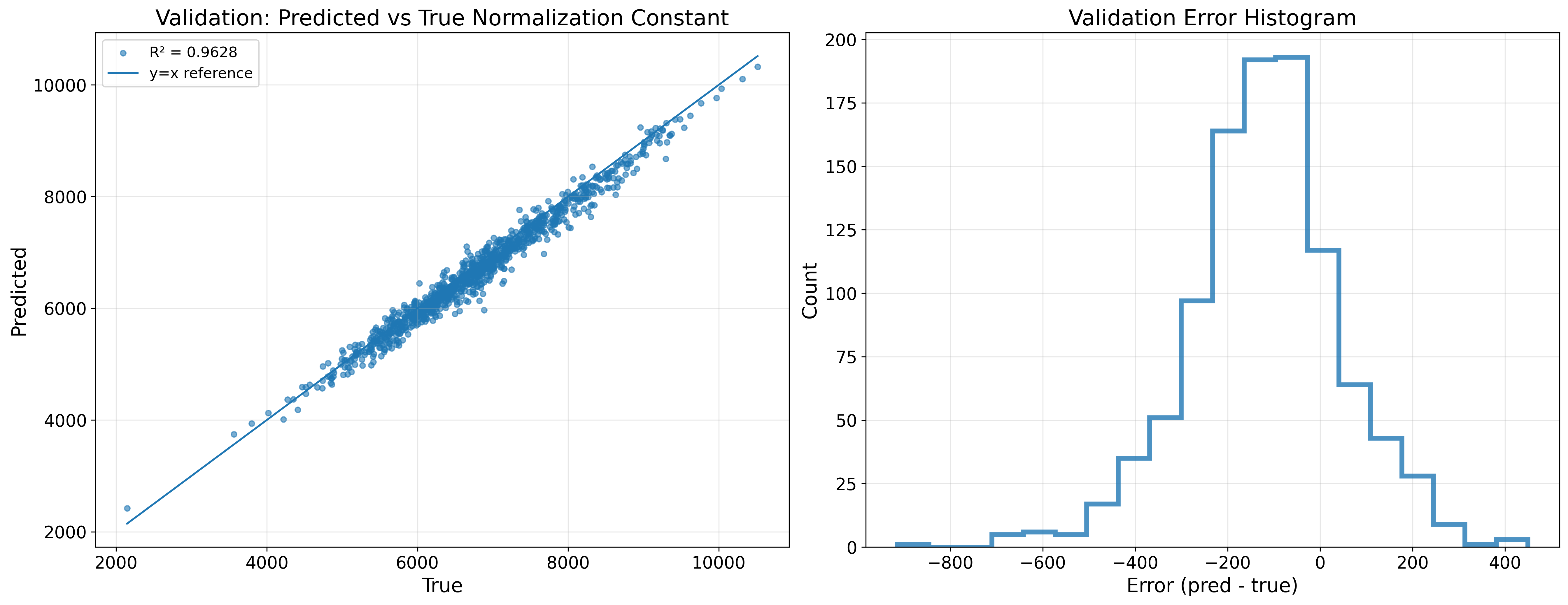}
    \caption{Validation performance for prediction of the spectral normalisation constant \(\alpha\), defined as the integrated discrete spectral weight within the selected frequency window. The left panel compares the predicted and reference values for molecules in the validation set. The solid line denotes ideal agreement, \(y=x\), and the coefficient of determination is \(R^{2}=0.9628\). The right panel shows the distribution of residual errors,
    defined as predicted minus reference. Most predictions lie close to the 
    reference line, although the residual distribution shows a small tendency
    towards negative errors, indicating slight underestimation for part of the
    validation set.}
\label{fig:normalisation_constant}

\end{figure}

The normalised spectral-shape target preserves the distribution of intensity over the investigated frequency interval but removes the overall spectral scale. We therefore performed an auxiliary regression experiment in which the spectral normalisation constant,
\begin{equation}
    \alpha
    =
    \sum_{\ell=1}^{L}s_{\ell},
\end{equation}
was predicted separately. Since the summation is restricted to the \(L=900\) bins considered in this work, \(\alpha\) represents the integrated spectral weight within the investigated interval of \(0\)--\(0.45\)~a.u., rather than the total oscillator strength over the complete excitation spectrum.

Figure~\ref{fig:normalisation_constant} compares the predicted and reference values on the validation set using the CNN model. The predictions follow the ideal \(y=x\) relation closely and achieve \(R^{2}=0.9628\), showing that the model captures most of the molecule-to-molecule variation in the overall spectral weight. The residual distribution is concentrated near zero, although its maximum is slightly shifted towards negative values. This indicates a modest tendency to underestimate the normalisation constant for part of the validation set, particularly towards the upper end of the reference-value range.

The result shows that the global spectral scale can be learned separately from the normalised line shape. In principle, the spectrum in its original intensity scale can therefore be reconstructed as
\begin{equation}
    \widehat{\mathbf{s}}
    =
    \widehat{\alpha}\,
    \widehat{\mathbf{p}},
\end{equation}
where \(\widehat{\mathbf{p}}\) is the predicted unit-normalised spectrum. Nevertheless, the principal comparisons in this work focus on \(\widehat{\mathbf{p}}\), because peak positions, relative peak intensities, and the redistribution of spectral weight are the features most directly relevant to spectral-shape comparison and molecular assignment. Prediction of \(\alpha\) is included as an auxiliary result demonstrating that the removed intensity scale is also recoverable with high accuracy. Due to the Thomas-Reiche-Kuhn sum rule (see Ref.\cite{U14}), the value of $\alpha$ (for the visible-UV spectra) is related to the normalization of the (valence) density (note it is not exactly determined due to the restriction of the frequency range), so it is not a surprise it can be well learned.

\subsection*{Dependence on chemical categories}

\begin{figure}[t]
    \centering
    \includegraphics[width=0.9\linewidth]{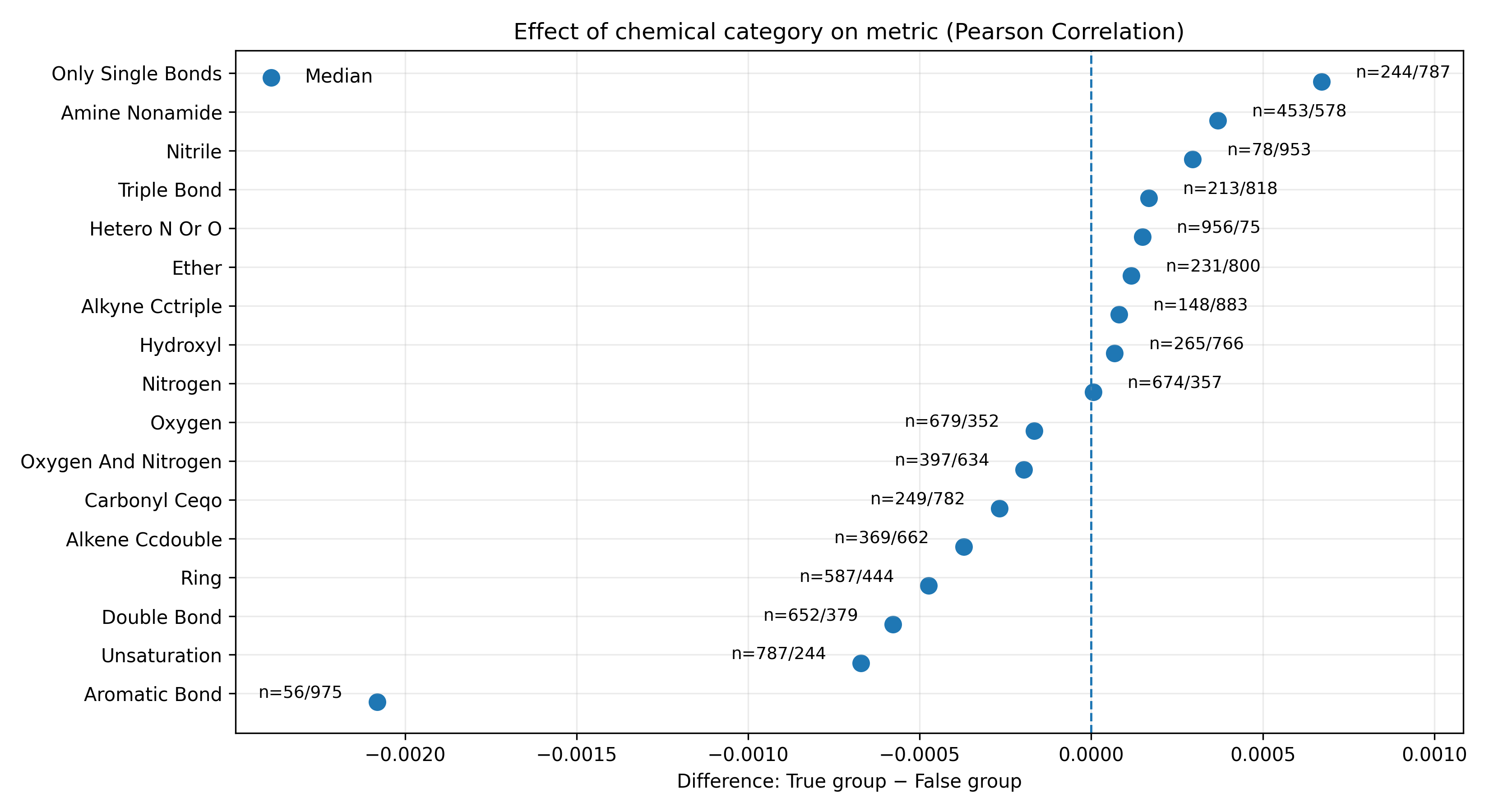}
    \caption{Effect of chemical categories on the median Pearson correlation between predicted and reference absorption spectra. Each point shows the difference in median Pearson correlation between molecules where the category is present and absent, computed as True group minus False group. Positive values indicate improved spectral-shape agreement for molecules containing the category, while negative values indicate reduced agreement. The annotations report the sample counts as $n_{\mathrm{true}}/n_{\mathrm{false}}$.}
    \label{fig:chem_delta_spectra_pearson}
\end{figure}

\begin{figure}[t]
    \centering
    \includegraphics[width=0.9\linewidth]{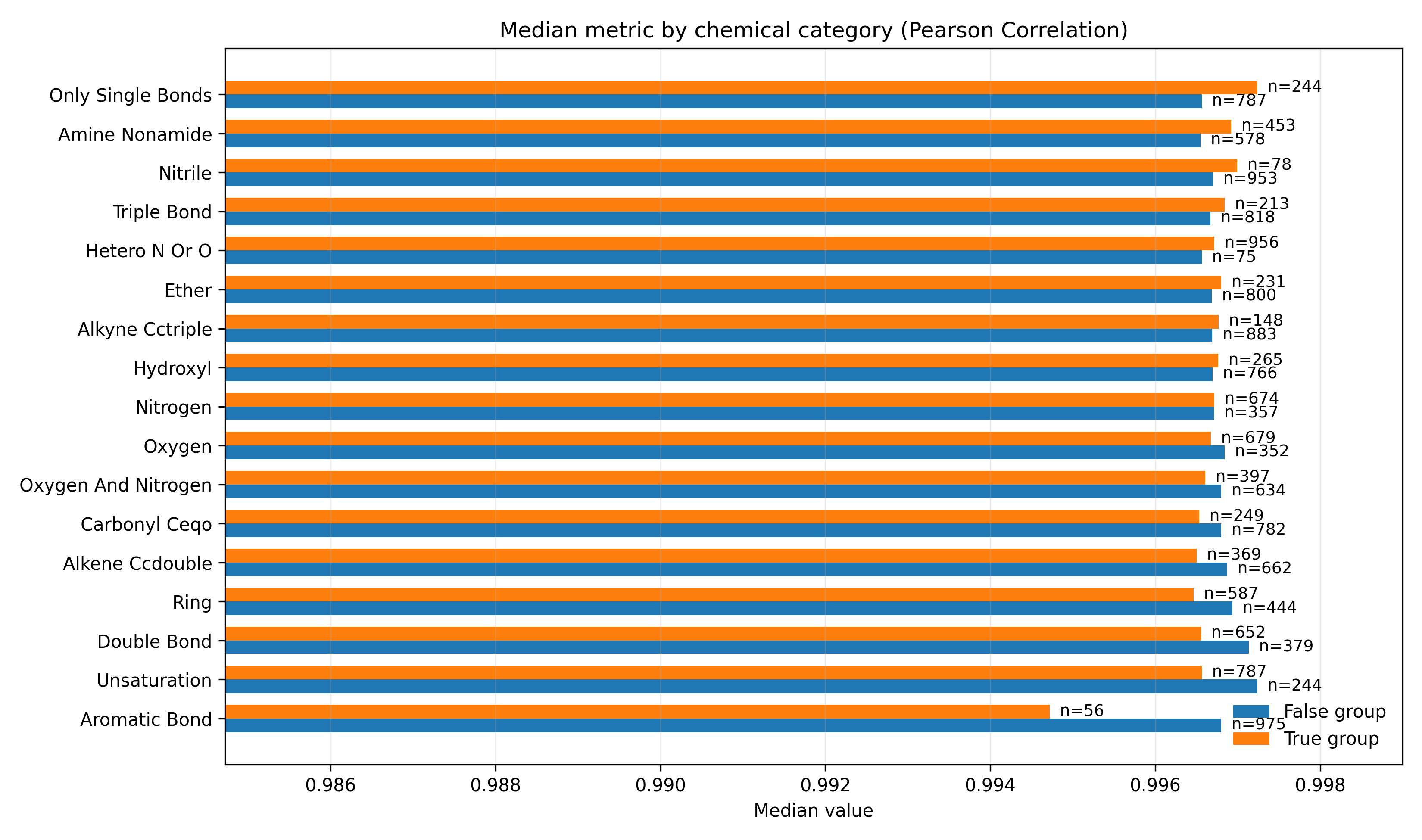}
    \caption{Median Pearson correlation between predicted and reference absorption spectra grouped by chemical category. For each category, the False group contains molecules where the category is absent, while the True group contains molecules where the category is present. The annotations indicate the number of molecules in each group. Higher values correspond to better agreement between the predicted and reference spectra.}
\label{fig:chem_true_false_spectra_pearson}
\end{figure}

Finally, we examine whether the density-CNN spectral performance changes systematically across different chemical categories. For absorption spectra, Fig.~\ref{fig:chem_true_false_spectra_pearson} shows the median Pearson correlation between predicted and reference spectra for molecules where each chemical category is absent or present. The corresponding difference plot in Fig.~\ref{fig:chem_delta_spectra_pearson} reports the true-minus-false change in median correlation.

The median correlations remain high across all major bonding and functional-group categories. In most cases, the difference between the true and false groups is only about $0.001$--$0.002$ in correlation, corresponding to roughly $0.1$--$0.2$ percentage points. This small variation indicates that the CNN is not performing well only for a narrow subset of molecules, but remains robust across diverse chemical environments.

\section*{Conclusions}

In this work, we investigated neural-network models for predicting molecular electronic properties associated with absorption spectra from two complementary molecular representations: three-dimensional electron-density fields and atom-centred molecular geometries. The models were evaluated on a task of computing the absorption spectrum.

For normalised full-spectrum prediction, the density-based 3D-CNN gives the strongest overall performance. The best CNN achieves a validation correlation of $0.9926$, a validation MAE of $7.14\times10^{-5}$, and a mean validation KL divergence of $6.66\times10^{-6}$. SchNet, DimeNet++, and MACE nevertheless provide strong geometry-based baselines, with validation correlations close to $0.98$ when trained using suitable KL or JS divergence objectives. The improvement obtained with the density representation is apparent across both pointwise and distributional measures. Relative to the best graph model, the CNN reduces the residual decorrelation, $1-r$, from $0.0205$ to $0.0074$, corresponding to a reduction of approximately $64\%$, while also reducing the spectral MAE by approximately a factor of two.

The computational requirements of the selected models are summarised in Tables~\ref{tab:model_compute} and~\ref{tab:inference_benchmark}. The density-based CNN requires substantially longer training and greater inference resources because it operates directly on large three-dimensional electron-density grids, in contrast to the more compact atom-centred representations used by the graph models. Despite this increased computational requirement, CNN inference remains on the millisecond timescale once the ground-state density is available. The timing results therefore quantify the additional computational requirements associated with processing the richer volumetric representation, while the improved spectral prediction supports the value of electron density as an input for this task.

The use of the ground-state electron density as an input representation is physically motivated by the foundations of density functional theory. 
In this sense, the density is not merely a geometric descriptor, but a field variable that encodes the electronic environment produced by the nuclei and by electron--electron interactions. This is particularly relevant for optical absorption, which is governed by the electronic response of the system to an external electromagnetic perturbation.
However, it will also be true for other spectra such as vibrational IR or Raman and X-ray absorption. It is also not restricted to DFT based methods and could be used for spectra calculated using other \textit{ab-initio} methods, such as Bethe-Salpeter, ADC,  or EOM-CC.
A density-based representation provides a natural intermediate level between nuclear geometry and the full response calculation: it already contains information about bonding, charge localization, delocalization, and local electronic environments that must otherwise be inferred indirectly from atomic coordinates and nuclear charges alone.

Although the ground-state density must first be computed before such a model can be applied, this step is typically less demanding than computing the absorption spectrum itself through explicit excited-state or time-dependent response calculations. The proposed approach therefore does not avoid electronic-structure theory entirely; rather, it replaces the more expensive density-to-spectrum response step with a learned surrogate. This makes the method physically informed, because the machine-learning model acts on a quantity that is central to electronic-structure theory, while also being computationally advantageous compared with direct TDDFT calculations of spectra.

The graph-model experiments remain important. SchNet and DimeNet++ achieve very similar performance across the best loss functions, while MACE reaches the same regime using an equivariant message-passing architecture. For MACE, the inclusion of vector equivariant features gives a small improvement over a scalar-only control, and the best results are obtained with relatively shallow message passing. The loss-function comparison shows that KL and JS divergence provide the best balanced performance for Softmax-normalized spectra, whereas direct Wasserstein optimization reduces the Wasserstein metric but does not give the best correlation or pointwise agreement.

The representative spectra show that the density-CNN captures dominant peaks, broad spectral profiles, and relative intensity changes between similar molecules. The remaining errors are concentrated in weak peaks, shoulders, and subtle fine-scale spectral features. The chemical-category analysis further shows that the CNN is robust across functional groups and bonding motifs: for most categories, the difference between molecules where the category is present or absent is only about $0.1$--$0.2$ percentage points in median spectral correlation. Larger shifts are observed for aromatic and other electronically distinctive groups, suggesting that targeted improvements in chemical balance and dataset diversity may further improve performance.

An important limitation of the present density-CNN model is that a conventional 3D-CNN is not intrinsically rotationally invariant, so predictions can depend on the orientation of the Cartesian density grid. The current results therefore demonstrate the predictive value of the electron-density representation within the orientation convention of the dataset, rather than complete rotational robustness. This limitation could be addressed through rotational data augmentation or rotation-aware architectures, together with complementary DFT-derived descriptors such as the electron localisation function (ELF), which provides additional information on electronic localisation and bonding.

Overall, these results demonstrate that machine-learning models can learn meaningful mappings from molecular representations to electronic spectra. For the normalised-spectrum task, the electron-density representation provides the strongest empirical performance among the models considered and is physically well motivated for absorption-spectrum prediction. Future work should focus on density-based and multimodal models that combine the direct electronic information available in the density field with the structural efficiency of graph representations. Such models may improve the prediction of fine spectral structure, rare chemical subgroups, rotational robustness, and uncertainty-calibrated spectral response functions.

\section*{Author contributions}

Siddharth Dhanpal: Methodology, Software, Formal analysis, Investigation, Validation, Visualization, Writing – original draft, Writing – review \& editing. Peter Elliott: Conceptualization, Methodology, Software, Data curation (TDDFT), Investigation, Writing – original draft, Writing – review \& editing. Paolo E. Trevisanutto: Methodology, Resources, Investigation, Writing – review \& editing. Alin Marin Elena: Methodology, Resources, Investigation, Writing – review \& editing. Gilberto Teobaldi: Conceptualization, Methodology, Resources, Investigation, Writing – review \& editing, Project administration, Supervision.

\section*{Conflicts of interest}

There are no conflicts to declare.

\section*{Data availability}

\noindent The dataset generated and analysed in this study is available from UKRI-STFC PSDI at \href{https://data-collections.psdi.ac.uk/records/pk3n6-s4778}{https://data-collections.psdi.ac.uk/records/pk3n6-s4778}.\\

\noindent The weights of ML model generated in this study will be available from eData, the STFC Research Data repository.\\ %

\noindent The code used for training and evaluating the machine-learning models is available from github under a MIT license at \href{https://github.com/stfc-ai4s/QM7-Absorption-ML}{https://github.com/stfc-ai4s/QM7-Absorption-ML}.\\

\section*{Acknowledgments}

 This work was supported by the Ada Lovelace Centre at the Science and Technology Facilities Council (https://adalovelacecentre.ac.uk/), the Physical Sciences Data Infrastructure (https://psdi.ac.uk; jointly STFC and the University of Southampton) under grants EP/X032663/1 and EP/X032701/1, and EPSRC under grants EP/W026775/1 and EP/V028537/1.

\bibliographystyle{apsrev4-2}
\bibliography{ML_TDDFT_article}
\end{document}

%% file: dft.tex
\def\bea{\begin{eqnarray}}
\def\eea{\end{eqnarray}}
\def\ben{\begin{equation}}
\def\een{\end{equation}}
\def\benu{\begin{enumerate}}
\def\enu{\end{enumerate}}
\def\bei{\begin{itemize}}
\def\eei{\end{itemize}}

\def\sss{\scriptscriptstyle\rm}

\def\1var{(\bx_1...\bx\N)}

\def\br{{\bf r}}

\def\bx{{\br t}}

\def\s{_{\sss S}}
\def\xc{_{\sss XC}}

\def\N{_{\sss N}}
\def\H{_{\sss H}}

\def\ext{_{\rm ext}}

\def\sph_int{ {\int d^3 r}}